\documentclass[preprint,aps,pra,showpacs,floatfix]{revtex4-1}
\usepackage[utf8]{inputenc}
\usepackage[OT1]{fontenc}
\usepackage{amsmath}
\usepackage{amsfonts}
\usepackage{amssymb}
\usepackage{bm}
\usepackage{graphicx}
\usepackage[left=2cm,right=2cm,top=2cm,bottom=2cm]{geometry}
\usepackage{longtable}
\allowdisplaybreaks

\newcommand{\veps}{\varepsilon}
\newcommand{\balpha}{\bm{\alpha}}

\newcommand{\br}{\bm{r}}
\newcommand{\bx}{\bm{x}}
\newcommand{\by}{\bm{y}}

\newcommand{\bnr}{\bm{\hat r}}
\newcommand{\dd}{{\rm d}}
\newcommand{\be}{\begin{eqnarray}}
\newcommand{\ee}{\end{eqnarray}}

\newcommand{\dvec}[2]{{\left( \begin{array}{c} #1 \\ #2 \\ \end{array} \right)}}

\newcommand{\ket}[1]{|#1\rangle}
\newcommand{\braket}[2]{\langle#1|#2\rangle}
\newcommand{\matrixel}[3]{\left\langle#1\left|\vphantom{#1}#2\vphantom{#3}\right|#3\right\rangle}

\begin{document}

\title{Ground-state ionization energies of boronlike ions}

\author{A.~V.~Malyshev$^{1,2}$, D.~A.~Glazov$^{1,3}$, A.~V.~Volotka$^{1,4}$, I.~I.~Tupitsyn$^{1,5}$,  
V.~M.~Shabaev$^{1}$, G.~Plunien$^{6}$, and Th.~St\"ohlker$^{4,7,8}$}

\affiliation{
$^1$ Department of Physics, St. Petersburg State University,
Universitetskaya 7/9, 199034 St.~Petersburg, Russia \\
$^2$ ITMO University, Kronverkskiy 49, 197101 St.~Petersburg, Russia \\
$^3$ SSC RF ITEP of NRC ``Kurchatov Institute'', B. Cheremushkinskaya 25, 117218 Moscow, Russia \\
$^4$ Helmholtz-Institut Jena, Fr\"obelstieg 3, D-07743 Jena, Germany \\
$^5$ Center for Advanced Studies, Peter the Great St. Petersburg Polytechnic University,
Polytekhnicheskaja 29, 195251 St.~Petersburg, Russia \\
$^6$ Institut f\"ur Theoretische Physik, Technische Universit\"at Dresden,
Mommsenstra{\ss}e 13, D-01062 Dresden, Germany \\
$^7$ GSI Helmholtzzentrum f\"ur Schwerionenforschung GmbH, Planckstra{\ss}e 1, D-64291 Darmstadt, Germany \\
$^8$ Institut f\"ur Optik und Quantenelektronik, Friedrich-Schiller-Universit\"at, Max-Wien-Platz 1, D-07743 Jena, Germany \\
}


\begin{abstract}

High-precision QED calculations of the ground-state ionization energies are performed
for all boronlike ions with the nuclear charge numbers in the range  $16 \leqslant Z\leqslant 96$.
The rigorous QED calculations are performed within the extended Furry picture
and include all many-electron QED effects up to the second order of the perturbation theory.
The contributions of the third- and higher-order electron-correlation effects
are accounted for within the Breit approximation. The nuclear recoil and nuclear polarization
effects are taken into account as well.
In comparison with the previous evaluations of the ground-state ionization energies of boronlike ions
the accuracy of the theoretical predictions has been improved significantly.

\end{abstract}


\maketitle


\section{Introduction \label{sec:0}}

Quantum electrodynamics (QED) is well known as a powerful tool to describe systems of
electrically charged particles interacting via the electromagnetic forces.
Since the beginning of 1950's when this theory was formulated
it had a great success in interpreting and predicting experimental results.
With a possibility to study heavy few-electron ions experimentally, which appeared
in the middle of 1980's \cite{Beyer:2003:book:eng},
these systems became the subject of extensive theoretical investigations,
for review see Refs.~\cite{Mohr:1998:227,Sapirstein:2008:25,Shabaev:2008:1220,Glazov:2011:71,Volotka:2013:636}
and references therein.
In contrast to light atoms and ions, where the binding-strength parameter of the nucleus $\alpha Z$
($\alpha$ is the fine structure constant and $Z$ is the nuclear charge number)
can be used along with $\alpha$ to evaluate
QED effects by perturbation theory, for high-$Z$ systems this parameter is not small
and, therefore, all the calculations must be performed without any expansion in $\alpha Z$.
On the other hand, since the number of electrons $N$ in highly charged ions
is much smaller than the nuclear charge number $Z$,
it is possible to calculate interelectronic-interaction effects using the perturbation theory in the parameter $1/Z$.
This approach leads to quantum electrodynamics in the Furry picture~\cite{Furry:1951:115}, where
the interaction of electrons with nucleus is treated to all orders in $\alpha Z$.
Due to rather simple electronic structure of heavy few-electron ions as compared to neutral atoms
the uncertainty of the electron-correlation corrections doesn't prevent one to probe QED effects in such systems.
Therefore, highly charged ions are ideal candidates
to examine new theoretical methods developed for description of bound-electron systems.


State-of-the-art QED calculations of the energy levels in highly charged ions include
all relevant contributions up to the second order in $\alpha$.
To date, the corresponding calculations have been performed for highly charged ions
with the number of electrons from one to five~
\cite{Yerokhin:2015:033103,Artemyev:2005:062104,Yerokhin:2001:032109,Kozhedub:2010:042513,Sapirstein:2011:012504,
Malyshev:2014:062517,Malyshev:2015:012514,Artemyev:2007:173004,Artemyev:2013:032518,Malyshev:2016:NIMB:accepted}.
High-precision measurements of the binding and transition energies which are sensitive to the second-order corrections
confirm predictions made by QED theory to a high level of accuracy~
\cite{Schweppe:1991:1434,Stoehlker:1993:2184,Beiersdorfer:1998:1944,Beiersdorfer:1998:3022,
Bosselmann:1999:1874,Stoehlker:2000:3109,Brandau:2003:073202,Draganic:2003:183001,
Gumberidze:2004:203004,Gumberidze:2005:223001,Beiersdorfer:2005:233003,Mackel:2011:143002,
Kubicek:2014:032508,Bernhardt:2015:144008,Beiersdorfer:2015:032514,Epp:2015:020502_R,Kraft-Bermuth:2017:055603}
and allow one to test bound-state QED in the strong-field regime.
In future, several new facilities are planned or already commenced aiming for
further improvement of the experimental accuracy.
Among them are the high-precision X-ray spectroscopy projects
FOCAL~\cite{Beyer:2015:144010} and maXs calorimeter~\cite{Hengstler:2015:014054} implementing at the CRYRING@ESR
facility in GSI~\cite{Lestinsky:2016:797} as well as the mass-spectrometry project PENTATRAP~\cite{Repp:2012:983,Roux:2012:997} in Heidelberg.
In view of this, further extension and improvement of \textit{ab initio} QED calculations are of great importance.

The present investigation is focused on five-electron boronlike ions.
There are many relativistic calculations of the excited state energies of boronlike ions,
see, e.g., Refs.~\cite{Safronova:1996:2850,Safronova:1998:183,Koc:2005:46,Rynkun:2012:481}.
However, the many-electron QED effects were considered within some one-electron or semiempirical approximations in these studies.
The rigorous QED evaluation of the $2p_{3/2}-2p_{1/2}$ fine-structure splitting in boronlike ions was performed only
in Refs.~\cite{Artemyev:2007:173004,Artemyev:2013:032518}.
The main goal of the present paper is to calculate the ionization energies of the $1s^2 2s^2 2p_{1/2}$ ground state
for the nuclear charge number in the wide range $16 \leqslant Z\leqslant 96$.
The numerical approach that we use is generally similar to the procedure
discussed in Ref.~\cite{Malyshev:2015:012514},
where the calculations of the ground-state ionization energies for berylliumlike ions have been performed.
The approach combines the first and second orders of the QED perturbation theory
with the higher-order electron-correlation contributions evaluated within the Breit approximation.
Taking into account the results of Ref.~\cite{Artemyev:2013:032518} one can easily obtain
the ionization energies for the first excited state $1s^2 2s^2 2p_{3/2}$ of boronlike ions,
for this one needs to add the transition energies of Ref.~\cite{Artemyev:2013:032518} to the the 
ground-state ionization energies presented in this work.


The paper is organized as follows. In Sec.~\ref{sec:1} we describe the procedure which was used to
evaluate the ionization energies. In Sec.~\ref{sec:2} our numerical results are presented and compared
with the theoretical predictions obtained in previous works.

The relativistic units ($\hbar = c = 1$) and the Heaviside charge unit ($\alpha = e^2/4\pi, e<0$)
are used throughout the paper.


\section{Basic formulas \label{sec:1}}

In highly charged ions the number of electrons $N$ is much smaller than the nuclear charge number~$Z$.
As a result, the interelectronic interaction is suppressed by the factor~$1/Z$ compared to the
interaction of the individual electrons with the Coulomb field of the nucleus.
Therefore, a good starting point for systematic description of these systems within QED 
is the Furry picture~\cite{Furry:1951:115}.
In this framework one starts from the assumption that one-electron wave functions obey the Dirac equation
\begin{equation}
\left[-i \balpha \cdot \nabla + \beta m + V ( \bm{r} )\right] \psi_n ( \bm{r} ) = \varepsilon_n \psi_n ( \bm{r} ),
\label{DirEq}
\end{equation}
where in the simplest case $V ( \bm{r} )$ is the potential of the nucleus~$V_{\rm{nucl}}( \bm{r} )$.
Alternatively, one can choose
the potential~$V ( \bm{r} )$ to be an effective potential, that is the sum of the nuclear potential
and some screening potential:
\begin{equation}
V( \bm{r} ) \rightarrow V_{\rm{eff}}( \bm{r} ) = V_{\rm{nucl}}( \bm{r} ) + V_{\rm{scr}}( \bm{r} ).
\label{EffPot}
\end{equation}
The screening potential is employed to take into account
the interelectronic-interaction effects partly from the very beginning.
In the present work we follow this alternative choice of the zeroth order approximation, which is known
as the extended Furry picture. This approach accelerates the convergence of the perturbation series
and was applied successfully to high-precision calculations of various atomic properties
\cite{Sapirstein:2002:042501,Sapirstein:2011:012504,Chen:2006:042510,Artemyev:2007:173004,
Yerokhin:2007:062501,Kozhedub:2010:042513,Artemyev:2013:032518,
Sapirstein:2003:022512,Sapirstein:2006:042513,Oreshkina:2007:889:note,
Kozhedub:2007:012511,Volotka:2008:062507,Glazov:2006:330,Volotka:2014:253004,Zubova:2016:052502}.
In addition, the use of the effective potential improves significantly
the numerical accuracy of the calculations, since it can remove the quasi-degeneracy
of the states with the same symmetry, which may take place if the pure Coulomb
potential is employed in the initial approximation.

We note here, that the zeroth-order results depend strongly on the choice of the effective potential.
If one could treat the QED and correlation effects
to all orders of the perturbation theory, the final results would be independent on screening potential employed.
In practice, however, state-of-the-art calculations for highly charged ions
are limited by the consideration of the second-order QED corrections.
The deviations of the final results obtained with the use of the different screening potentials can provide
an estimation of the uncalculated higher-order contributions.

In the present work we employ several different types of the screening potential.
The first one is the core-Hartree (CH) potential induced by the $1s^2 2s^2$ core.
This potential can be constructed from the radial charge density of the $1s$ and $2s$ electrons:
\begin{gather}
V_{\rm CH}(r) = \alpha \int_0^\infty d r' \frac{1}{r_>} \rho_{\rm CH} (r'),  \label{CHpot} \\[1mm]
\rho_{\rm CH}(r) = 2 \sum_{c=1s,2s} \left[ G_{c}^2(r) + F_{c}^2(r)  \right], \qquad
\int_0^\infty \rho_{\rm CH}(r) dr = N-1,    \label{CHdens}
\end{gather}
where $N=5$ is the total number of the electrons, $r_>={\rm max}(r,r')$,
$G/r$ and $F/r$ are the large and small radial components of the Dirac wave function
\begin{equation}
\psi(\br) = \frac{1}{r} \dvec{ G(r) \Omega_{\kappa m} (\bnr) }{ i F(r) \Omega_{-\kappa m} (\bnr) },
\label{eq:wf}
\end{equation}
$\Omega_{\kappa m}$ is the spin-angular spinor, and $\bnr = \br / r$.

The next three potentials are built for the $1s^2 2s^2 2p_{1/2}$ electron configuration.
The local Dirac-Fock~(LDF) potential is generated from the wave functions evaluated within
the Dirac-Fock approximation \cite{pot:LDF} and reproduces the energies and wave functions of the
$2p_{1/2}$ state at the corresponding level.
The Kohn-Sham~(KS) and Perdew-Zunger~(PZ) potentials are
formulated in the framework of the density-functional theory.
The KS potential is given by the following expression~\cite{pot:KS}:
\begin{equation}
V_{\rm{KS}}(r) = \alpha \int_0^\infty d r' \frac{1}{r_>} \rho_t(r')
- \frac{2}{3} \frac{\alpha}{r} \left( \frac{81}{32\pi^2}  r \rho_t(r) \right)^{1/3},\label{KSpot}
\end{equation}
where $\rho_t$ is the radial density of all electrons:
\begin{equation}
\rho_t(r) = \rho_{\rm CH}(r) + \left[ G_{2p_{1/2}}^2(r) + F_{2p_{1/2}}^2(r) \right], \qquad
\int_0^\infty \rho_{t}(r) dr = N.    \label{KSdens}
\end{equation}
The second term in Eq.~(\ref{KSpot}) describes the exchange part of the interelectronic interaction.
We use the Latter correction~\cite{pot:Latter} to restore
the proper asymptotic behavior of the KS potential at large distances.
The PZ potential contains an additional correlation term in comparison to the KS potential~\cite{pot:PZ}.
The self-interaction correction for the PZ potential
is taken into account according to Ref.~\cite{pot:PZ}.
In order to expand the range of the initial approximations used to evaluate ionization energies of boronlike ions
we consider the second PZ potential generated for the $1s^2 2s^2 2p_{3/2}$ electron configuration.
To distinguish those two kinds of the PZ potential in what follows we will add the indices 1 and 3
to label the potentials constructed for the configurations with the $2p_{1/2}$ and $2p_{3/2}$ electrons, respectively.
Therefore, the total number of the screening potentials used in the present work is equal to five.
We note, that at large distances $r$ all the screening potentials behave like $N_{\rm scr}\alpha/r$,
where $N_{\rm scr}=N-1$.

Solving Eq.~(\ref{DirEq}) with an effective potential~(\ref{EffPot}) for the state $2p_{1/2}$
one can obtain the zeroth-order approximation for the ionization energy
$E^{(0)}_{\rm Dirac} = \varepsilon_{2p_{1/2}} - mc^2$.
The QED and electron-correlation contributions have to be taken into account
by a perturbation theory. To derive formal expressions for the terms of the perturbation series we use
the two-time Green function (TTGF) method~\cite{TTGF}.
It should be noted that within the extended Furry picture the potential
$\delta V( \bm{r} ) = -V_{\rm{scr}}( \bm{r} )$ is to be added into the QED interaction Hamiltonian.
The counterterm $\delta V( \bm{r} )$ must be accounted for perturbatively in order to
avoid the double counting of the screening effects.
In order to perform bound-state QED calculations
one needs a quasicomplete basis set of the one-electron solutions of the Dirac equation
to have a spectral representation of the Dirac-Coulomb Green function
\begin{equation}
G(\omega, \bm{r}_1,\bm{r}_2) = \sum_n \frac{\psi_n(\bm{r}_1)\psi^\dagger_n(\bm{r}_2)}{\omega - \varepsilon_n(1-i0)}.
\label{GreenF}
\end{equation}
The basis wave functions were constructed from the B-splines~\cite{splines:B}
with the use of the dual kinetic balance (DKB) approach~\cite{splines:DKB}.

\begin{figure}
\begin{center}
\includegraphics[width=12cm]{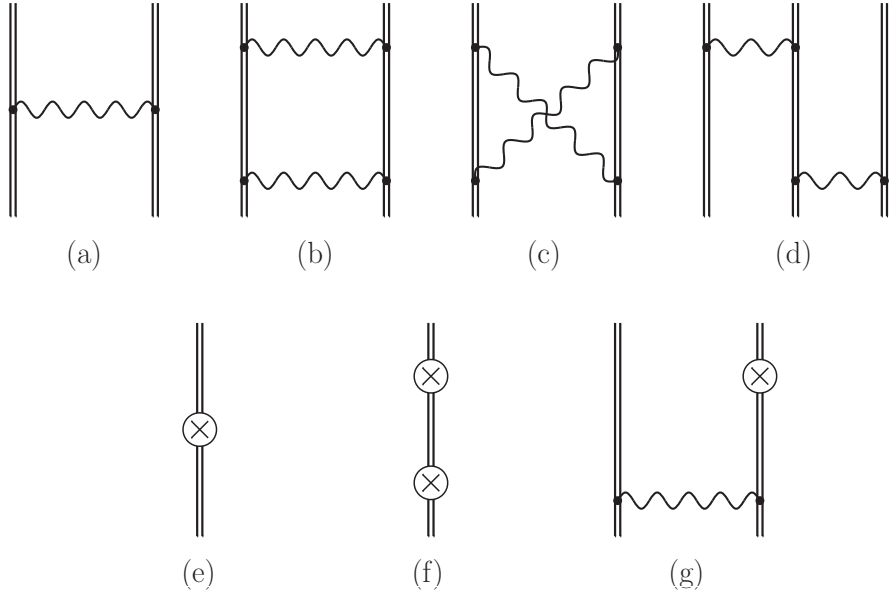}
\caption{\label{fig:int}
The interelectronic-interaction diagrams in the extended Furry picture.
The double line denotes the electron propagator in the effective potential (\ref{EffPot}).
The wavy line corresponds to the photon propagator.
The symbol $\otimes$ represents the local screening potential counterterm $\delta V( \bm{r} ) = -V_{\rm{scr}}( \bm{r} )$.}
\end{center}
\end{figure}

The evaluation of the ionization energies of boronlike ions can be accomplished in several steps.
At first, we will discuss the calculation of the interelectronic-interaction corrections.
As it was mentioned above, one can account for the electron-electron interaction partly
replacing the potential of the nucleus in Eq.~(\ref{DirEq}) with some effective potential. The remaining part of the
interelectronic interaction is to be considered within the perturbation theory.
In Fig.~\ref{fig:int} the set of the first- and second-order Feynman diagrams describing
the interelectronic interaction is shown. The double line corresponds to the electron
propagator in the effective potential~(\ref{EffPot}). The wavy line represents the photon propagator.
Finally, the circle with a cross depicts the counterterm $\delta V( \bm{r} )$.
The ionization energy of the $2p_{1/2}$ electron can be obtained by subtracting
the binding energy of the Be-like ion from the binding energy of the corresponding B-like ion.
The diagrams with only core electrons as incoming (outgoing) electron lines
cancels each other in this difference and don't contribute to the ionization energy under consideration.
Therefore, in what follows we will work only with the diagrams where one of the incoming (outgoing) electron lines
corresponds to the $2p_{1/2}$ valence electron.
The formulas for the contributions of the first- and second-order diagrams shown in Fig.~\ref{fig:int}
can be derived with the use of the TTGF method. The formal expressions for all the relevant contributions
are presented in Appendix~\ref{app:A}.

At the next stage we have to evaluate the interelectronic-interaction contributions of the third and higher orders,
which are also important. In the present work we have calculated
these corrections within the Breit approximation using two different approaches.
First of all, as in Refs.~\cite{Malyshev:2014:062517,Malyshev:2015:012514}, we have employed the 
configuration-interaction Dirac-Fock-Sturm (CI-DFS) method \cite{CI,CI2} to solve
the Dirac-Coulomb-Breit (DCB) equation yielding the total binding energy.
In order to separate the desired third- and higher-order contributions from the total
CI-DFS results the following procedure has been used. A free parameter $\lambda$ is introduced
into the DCB Hamiltonian
\begin{gather}
H_{\rm DCB}(\lambda) = \Lambda^{(+)} \Big[ H_0 + \lambda V_{\rm int} \Big] \Lambda^{(+)} ,
		\label{eq:H_lambda_DCB} \\[1mm]
H_0 = \sum_i \left[ \balpha_i \cdot \bm{p}_i + \beta_i m + V_{\rm eff}^{(i)} \right],
		\label{eq:H0_DCB} \\[1mm]
V_{\rm int} = \sum_{i<j} \left[ V_{\rm C}^{(i,j)} + V_{\rm B}^{(i,j)} \right] - \sum_i V_{\rm scr}^{(i)},
		\label{eq:V_DCB} \\[1mm]
V_{\rm C}^{(i,j)} =  \frac{\alpha}{r_{ij}}, \qquad
V_{\rm B}^{(i,j)} = -\frac{\alpha}{2} \left[ \frac{ \balpha_i \cdot \balpha_j }{ r_{ij} }
                          + \frac{(\balpha \cdot \bm{r}_{ij})(\balpha_j \cdot \bm{r}_{ij})}{r_{ij}^3}  \right],
        \label{eq:V_CB_DCB}
\end{gather}
where $\Lambda^{(+)}$ is the product of the one-electron projectors on the positive-energy states
(which correspond to the potential $V_{\rm eff}$),
$H_0$ is the unperturbed Hamiltonian, $V_{\rm int}$ describes the interaction of the electrons with each other and
with the counterterm potential $-V_{\rm scr}$, $V_{\rm C}$ and $V_{\rm B}$ are the Coulomb and Breit parts of
the electron-electron interaction operator in the Breit approximation, respectively.
The Hamiltonian~(\ref{eq:H_lambda_DCB}) coincides with the original DCB Hamiltonian for $\lambda=1$.
The energy $E$ evaluated within the CI-DFS method becomes a function of the parameter $\lambda$
when the Hamiltonian~(\ref{eq:H_lambda_DCB}) is used in the calculations.
One can expand the energy $E(\lambda)$ in powers of $\lambda$
\begin{gather}
E(\lambda) = E^{(0)} + \lambda E^{(1)} + \lambda^2 E^{(2)} + \sum_k^\infty \lambda^k E^{(k)},
		\label{eq:E_expansion} \\[1mm]
E^{(k)} = \left. \frac{1}{k!} \frac{\dd^k}{\dd \lambda^k} E(\lambda) \right |_{\lambda=0}.
		\label{eq:E_expansion_k}
\end{gather}
It can be seen that the coefficients $E^{(k)}$ correspond to the different orders of the perturbation theory.
Therefore, for the contribution of the third and higher orders
\begin{equation}
E^{(\geqslant 3)} = \sum_{k=3}^\infty  E^{(k)}
\label{eq:E_ho_def}
\end{equation}
one can obtain the following expression in the framework of the CI-DFS approach
\begin{equation}
E^{(\geqslant 3)} = \left. E(\lambda) \right|_{\lambda=1} - E^{(0)} - E^{(1)} - E^{(2)},
\label{eq:E_ho}
\end{equation}
where the coefficient $E^{(0)}$, $E^{(1)}$, and $E^{(2)}$ have to be determined numerically
according to Eq.~(\ref{eq:E_expansion_k}).

Along with the CI-DFS calculations, in the present work the higher-order interelectronic-interaction corrections have been evaluated also with the use of the recursive formulation of the perturbation theory.
The detailed description of this approach is presented in our recent work~\cite{Glazov:2016:NIMB:accepted}. 
Below we consider its basic principles.

While the perturbation expansion for the energy is given by Eq.~(\ref{eq:E_expansion}) with $\lambda=1$, we have the following expansion for the many-electron wave function $\ket{A}$ of the state under consideration
\begin{align}
\label{eq:ta}
  \ket{A}
   &= \sum_{k=0}^{\infty} \ket{A^{(k)}}
    = \sum_{k=0}^{\infty} \sum_N \ket{N^{(0)}} \braket{N^{(0)}}{A^{(k)}}
\,.
\end{align}
The finite basis set of the many-electron wave functions $\ket{N^{(0)}}$ consists of the Slater determinants, being the eigenfunctions of $H_0$ with the eigenvalues $E_N^{(0)}$,
\begin{equation}
\label{eq:A_0}
  \Lambda^{(+)} H_0 \Lambda^{(+)} \ket{N^{(0)}} = E_N^{(0)} \ket{N^{(0)}}
\;.
\end{equation}
The Slater determinants are made of the one-electron wave functions constructed within the DKB method \cite{splines:DKB}.  It provides the same zeroth-order approximation that is used for the calculation of the first- and second-order QED contributions assuming the same choice of the screening potential. The energy corrections $E^{(k)}$ and the coefficients $\braket{N^{(0)}}{A^{(k)}}$ can be found via the recursive system of equations,
\begin{align}
\label{eq:rec-en}
  E^{(k)} =& \sum_M \matrixel{A^{(0)}}{V_{\rm int}}{M^{(0)}} \braket{M^{(0)}}{A^{(k-1)}}
    - \sum_{j=1}^{k-1} E^{(j)} \braket{A^{(0)}}{A^{(k-j)}}
\,,\\
\label{eq:rec-na}
  {\left. \braket{N^{(0)}}{A^{(k)}} \right|}_{N \neq A} =\,& \frac{1}{E^{(0)} - E_N^{(0)}}
   \left[ \sum_M \matrixel{N^{(0)}}{V_{\rm int}}{M^{(0)}} \braket{M^{(0)}}{A^{(k-1)}}
    - \sum_{j=1}^{k-1} E^{(j)} \braket{N^{(0)}}{A^{(k-j)}} \right]
\,,\\
\label{eq:rec-aa}
  \braket{A^{(0)}}{A^{(k)}} =\,& - \frac{1}{2}\,\sum_{j=1}^{k-1}
    \sum_M \braket{A^{(j)}}{M^{(0)}} \braket{M^{(0)}}{A^{(k-j)}}
\,,
\end{align}
with the initial values,
\begin{align}
  {\left. \braket{N^{(0)}}{A^{(0)}} \right|}_{N \neq A} = 0
\,,
\;\;\;\;
  \braket{A^{(0)}}{A^{(0)}} = 1
\,.
\end{align}
The matrix elements of $V_{\rm int}$ with the Slater determinants are reduced to the one- and two-electron matrix elements according to the well-known combinatorial formulas. Each step of the recursion comprises only one-fold summation over the basis set for each $\braket{N^{(0)}}{A^{(k)}}$, in contrast to the conventional form of the perturbation theory. It makes the procedure computationally efficient and allows one to access essentially arbitrary high order of the perturbation theory. The desired contribution of the third- and higher-order interelectronic-interaction effects $E^{(\geqslant 3)}$ can be obtained by the direct summation in Eq.~(\ref{eq:E_ho_def}). The summation is terminated according to the total accuracy that one aims at. The energies obtained by means of the CI-DFS method and the direct summation of the perturbation series are found to be in a good agreement. The application of the fully independent approaches based on the completely different grounds allows one to perform more reliable estimation of the accuracy of the calculations.

\begin{figure}
\begin{center}
\includegraphics[width=13cm]{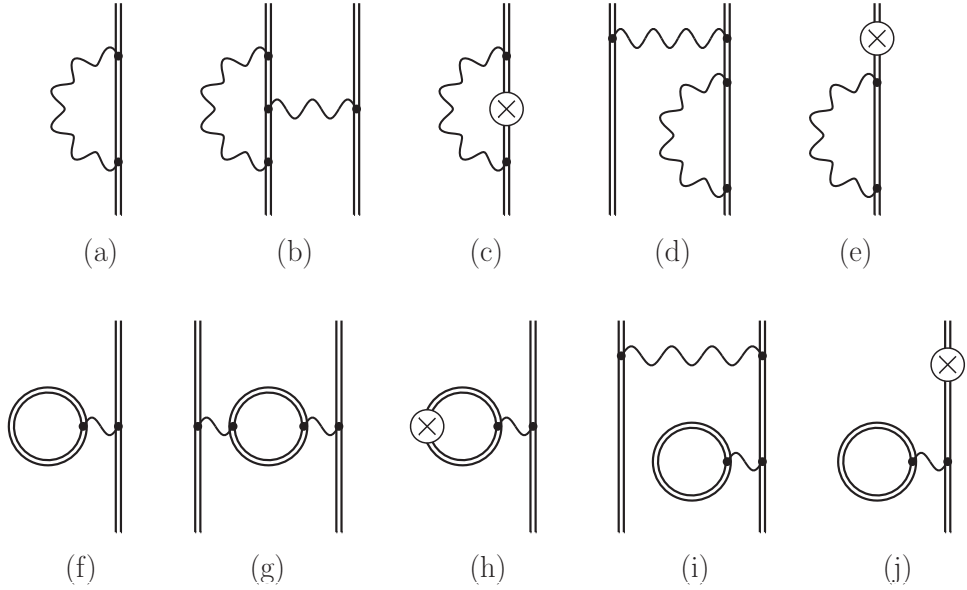}
\caption{\label{fig:qed}
First- and second-order QED diagrams (excluding the one-electron two-loop diagrams)
in the extended Furry picture.
The notations are the same as in Fig.~\ref{fig:int}.}
\end{center}
\end{figure}

At the next step one has to account for the radiative corrections.
All the necessary one-electron and many-electron one-loop Feynman diagrams are shown in Fig.~\ref{fig:qed}.
The first line in Fig.~\ref{fig:qed} presents the self-energy~(SE) diagrams, in the second
line the vacuum-polarization~(VP) diagrams are given.
The formal expressions for the corresponding corrections in the extended Furry picture
can be readily obtained in the framework of the TTGF method.
For convenience, all the formulas needed for evaluation of the ionization energies of boronlike ions
are collected in Appendix~\ref{app:B}.

We note, that in the present work in order to calculate
the first-order self-energy correction for the $2p_{1/2}$ electron
moving in the effective potential~(\ref{EffPot}) we employ the procedure
described in Refs.~\cite{Artemyev:2007:173004,Artemyev:2013:032518}.
The approach suggested there is a modification of the standard scheme 
(see Ref.~\cite{Artemyev:2013:032518} and references therein)
for the evaluation of the SE contributions.
According to the standard scheme one has to expand the
Dirac-Coulomb Green function in powers of the potential $V_{\rm{eff}}$
(see Eq.~(\ref{eq:SE_sigma}) in Appendix~\ref{app:B}), and this expansion
gives rise to the zero-, one-, and many-potential terms of the total SE correction.
The zero- and one-potential terms are both divergent and should be renormalized.
The renormalization procedure is discussed, e.g., in Ref.~\cite{Yerokhin:1999:800},
and after its application the calculation of the zero- and one-potential terms is straightforward.
The many-potential term corresponds to the part with two or more potentials
in the expansion of the Dirac-Coulomb Green function, and its evaluation is the most time consuming.
The calculation of the many-potential term for the $2p_j$ states in low-$Z$ ions in the case of the extended Furry picture
is complicated because of the slow convergence of the partial-wave expansion for this contribution,
see Ref.~\cite{Artemyev:2013:032518} for details.
In Refs.~\cite{Artemyev:2007:173004,Artemyev:2013:032518} it was suggested to separate
the part of the many-potential term that defines the slow convergence,
namely the two-potential term, that is the contribution with two potentials
in the Green function expansion. The separated two-potential term can be evaluated to a high accuracy
using the analytical representation for the free-electron propagator.
After the evaluation of the two-potential term is accomplished
the calculation of the remaining part of the many-potential term does not cause any difficulties.

In order to control the total accuracy of the one-electron SE contribution,
we have also performed the calculations employing the approach described, e.g., in Ref.~\cite{Kozhedub:2010:042513}
in addition to the method of Refs.~\cite{Artemyev:2007:173004,Artemyev:2013:032518}.
By means of the standard procedure~\cite{Yerokhin:1999:800} we have evaluated the screening effect
to the first-order SE correction, that is the difference between this contribution
calculated with and without the screening potential.
By adding the self-energy correction evaluated for the electron in the Coulomb field~\cite{Yerokhin:2015:033103},
we have obtained the results which are in a good agreement with the values calculated by the first approach.

Next, we have to take into account the contributions corresponding to the two-loop one-electron diagrams.
The consideration of these corrections completes the rigorous QED treatment of the ground-state ionization energies
of boronlike ions to the second order in $\alpha$ within the Furry picture.
The nonperturbative in $\alpha Z$ evaluation of the two-loop diagrams is a very difficult problem,
which is not fully accomplished yet. The most significant achievement in this field is related to
the calculation of the two-loop self-energy  correction by Yerokhin \textit{et al.}
\cite{Yerokhin:2003:073001,Yerokhin:2006:253004,Yerokhin:2007:521,Yerokhin:2009:040501_R,Yerokhin:2010:57}.
The contributions of the diagrams with closed fermion loops have been considered in
Refs.~\cite{Mallampalli:1996:2714,Persson:1996:2805,Beier:1997:2761,Plunien:1998:177,Yerokhin:2008:062510},
the free-loop approximation has been used in a part of these terms.
In the present work to account for the two-loop one-electron corrections we use the results from~
Refs.~\cite{Yerokhin:2008:062510,Yerokhin:2015:033103}.

In all the contributions discussed so far the nucleus was considered as a motionless source of the
external electrical field, i.e., it was assumed that the nucleus has the infinite mass.
The high-precision calculations of highly-charged ions have to go beyond this
external field approximation and account for the nuclear recoil corrections.
The full relativistic theory of the nuclear recoil effect to the first order in $m/M$ ($M$ is the nuclear mass)
and to all orders in $\alpha Z$ can be formulated only within QED
\cite{Shabaev:1985:394:note,Shabaev:1988:107:note,Shabaev:1998:59,Adkins:2007:042508}.
The lowest-order relativistic nuclear recoil corrections can be evaluated by averaging the operator~
\cite{Shabaev:1985:394:note,Shabaev:1988:107:note,Palmer:1987:5987}
\begin{equation}
H_{M}=\frac{1}{2M} \sum_{i,j} \left\{ \bm{p}_i \cdot \bm{p}_j  - \frac{\alpha Z}{r_i} \left[ \bm{\alpha}_i + \frac{(\bm{\alpha}_i \cdot \bm{r}_i) \bm{r}_i }{r_i^2} \right] \cdot \bm{p}_j \right\}.
		\label{eq:BrRecoil}
\end{equation}
In the present work we calculate the expectation values of the operator~(\ref{eq:BrRecoil})
with the many-electron wave functions obtained in the framework of the CI-DFS method.
Such an approach allows one to take into account the nuclear recoil effect within the
Breit approximation to all orders in $1/Z$.
This correction has been evaluated also with the wave functions obtained
within the recursive perturbation theory (\ref{eq:ta}). The contribution of the $k$-th order in $1/Z$ is given by,
\begin{align}
  \Delta E_{\rm rec,Breit}^{(k)} =& \sum_{j=0}^{k} \sum_{M,N} \braket{A^{(j)}}{M^{(0)}} \matrixel{M^{(0)}}{H_{M}}{N^{(0)}} \braket{N^{(0)}}{A^{(k-j)}}
\,,
\end{align}
where the coefficients $\braket{A^{(j)}}{M^{(0)}}\equiv\braket{M^{(0)}}{A^{(j)}}$ are evaluated through Eqs.~(\ref{eq:rec-en}), (\ref{eq:rec-na}), and (\ref{eq:rec-aa}).
The results of both calculations were found to be in a good agreement with each other.

The nuclear recoil corrections which are beyond the Breit approximation are referred to as the
QED nuclear recoil effects. We have evaluated these contributions to the zeroth order in $1/Z$.
The one-electron part of the QED nuclear recoil corrections in the case of the point nucleus can be expressed as follows~
\cite{Shabaev:1985:394:note,Shabaev:1988:107:note}:
\begin{eqnarray}
\Delta E_{\rm 1el}^{\rm rec} &=& \frac{i}{2\pi M} \int_{-\infty}^{\infty}\! d\omega \,
\left\langle v \left| \left( {\bm D}(\omega)-\frac{[{\bm p},V]}{\omega+i0} \right) \right. \right. \nonumber \\[1mm]
&& \times G(\omega + \varepsilon_v) \left. \left. \left( {\bm D}(\omega)+\frac{[{\bm p},V]}{\omega+i0} \right)
\right| v  \right\rangle,
		\label{QedRecoil_1el}
\end{eqnarray}
where ${\bm p}$ is the momentum operator, $D_m(\omega)=-4\pi \alpha Z \alpha_l D_{lm}(\omega)$,
and $D_{lm}$ is the transverse part of the photon propagator considered in the Coulomb gauge.
The two-electron part has the form~\cite{Shabaev:1988:107:note}
\begin{eqnarray}
\Delta E_{\rm 2el}^{\rm rec} &=&
  \frac{1}{M} \sum_{c=1s,2s} \sum_{\mu_c}
  \Big\{ \langle v | {\bm p} | c \rangle \langle c | {\bm D}(\Delta) - {\bm D}(0) | v \rangle  \nonumber \\[1mm]
  &&   + \langle v | {\bm D}(\Delta) - {\bm D}(0) | c \rangle \langle c | {\bm p} | v \rangle
       - \langle v | {\bm D}(\Delta) | c \rangle \langle c | {\bm D}(\Delta) | v \rangle \Big\},
		\label{QedRecoil_2el}
\end{eqnarray}
where $\Delta=\veps_v-\veps_c$.
One can partly take into account the nuclear size corrections to the QED nuclear recoil contributions~(\ref{QedRecoil_1el})
and (\ref{QedRecoil_2el}) by replacing the potential $V=-\alpha Z/r$, the Green function $G$,
the Dirac energies and wave functions of the point nucleus with the corresponding quantities
for the extended nucleus~\cite{Shabaev:1998:59}. For the $1s$ and $2s$ states in hydrogenlike ions
such calculations were performed in Refs.~\cite{Shabaev:1998:4235,Shabaev:1999:493},
see also Ref.~\cite{Aleksandrov:2015:144004}.
In the present paper, as in Refs.~\cite{Orts:2006:103002,Malyshev:2015:012514,Zubova:2016:052502},
we have evaluated the QED nuclear recoil corrections~(\ref{QedRecoil_1el})
and (\ref{QedRecoil_2el}) for the effective potential~(\ref{EffPot}) to account for the screening effects partly.

Finally, we have to consider the nuclear polarization corrections for high-$Z$ ions.
These contributions result from the electron-nucleus interactions that include
the excited intermediate nuclear states. We have used the data from
Refs.~\cite{Plunien:1991:5853,Plunien:1995:1119:note,Nefiodov:1996:227,Nefiodov:2002:081802,Volotka:2014:023002}
and the prescriptions from Ref.~\cite{Yerokhin:2015:033103} to take into account
the nuclear polarization effects.

Summarizing the description of the numerical approach which we use to evaluate the
ground-state ionization energies in boronlike ions it is worth noting that
the calculations of the many-electron QED corrections in Figs.~\ref{fig:int} and \ref{fig:qed}
have been carried out using both Feynman and Coulomb gauges for the electron-electron interaction propagator. 
The results of both calculations are in a good agreement with each other.
To describe the nuclear charge distribution we use the Fermi model with a thickness parameter equaled to 2.3~fm.
The nuclear radii are taken from Ref.~\cite{Angeli:2013:69} for the most abundant isotopes.
In the cases of $Z$ for which the data are absent in Ref.~\cite{Angeli:2013:69} we
have used the root-mean-square radii obtained by the approximate formula from Ref.~\cite{Johnson:1985:405}
with a $1\%$~uncertainty prescribed.


\section{Numerical results and discussions \label{sec:2}}

\begin{table}
\caption{Interelectronic-interaction contributions of the 3rd and higher orders to the ionization energies (in eV) of the ground state in boronlike calcium, xenon, and uranium with the LDF and PZ3 potentials. The results of the CI-DFS calculations are given for comparison.
\label{table:PT}}
\vspace{0.5cm}
\begin{tabular}{c@{\qquad}r@{}lr@{}l@{\qquad}r@{}lr@{}l@{\qquad}r@{}lr@{}l}
\hline
\hline
& \multicolumn{4}{c}{$Z=20$}
& \multicolumn{4}{c}{$Z=54$}
& \multicolumn{4}{c}{$Z=92$}\\
\cline{2-13}
PT order
& \multicolumn{2}{c}{LDF}
& \multicolumn{2}{c}{PZ3}
& \multicolumn{2}{c}{LDF}
& \multicolumn{2}{c}{PZ3}
& \multicolumn{2}{c}{LDF}
& \multicolumn{2}{c}{PZ3}\\
\hline
3     &          0.&3401   &          0.&4133   &       $-$0.&9392   &       $-$0.&4625   &       $-$3.&0172   &       $-$2.&4786   \\
4     &       $-$0.&5794   &       $-$0.&5313   &       $-$0.&1667   &       $-$0.&1973   &          0.&4857   &          0.&3527   \\
5     &          0.&1708   &          0.&1260   &          0.&0950   &          0.&0494   &          0.&0058   &          0.&0133   \\
6     &          0.&0544   &          0.&0564   &       $-$0.&0065   &          0.&0055   &       $-$0.&0356   &       $-$0.&0244   \\
7     &       $-$0.&0581   &       $-$0.&0436   &       $-$0.&0062   &       $-$0.&0036   &          0.&0129   &          0.&0072   \\
8     &          0.&0102   &          0.&0032   &          0.&0017   &       $-$0.&0000   &       $-$0.&0019   &       $-$0.&0007   \\
9     &          0.&0090   &          0.&0077   &          0.&0003   &          0.&0003   &       $-$0.&0005   &       $-$0.&0003   \\
10    &       $-$0.&0056   &       $-$0.&0030   &       $-$0.&0002   &       $-$0.&0000   &          0.&0004   &          0.&0002   \\
11    &       $-$0.&0001   &       $-$0.&0007   &          0.&0000   &       $-$0.&0000   &       $-$0.&0001   &       $-$0.&0000   \\
12    &          0.&0013   &          0.&0008   &          0.&0000   &          0.&0000   &       $-$0.&0000   &       $-$0.&0000   \\
13    &       $-$0.&0004   &       $-$0.&0001   &       $-$0.&0000   &          0.&0000   &          0.&0000   &          0.&0000   \\
14    &       $-$0.&0002   &       $-$0.&0001   &       $-$0.&0000   &       $-$0.&0000   &          0.&0000   &       $-$0.&0000   \\
15    &          0.&0001   &          0.&0000   &          0.&0000   &          0.&0000   &       $-$0.&0000   &          0.&0000   \\
\hline
3--$\infty$
      &       $-$0.&0579   &          0.&0287   &       $-$1.&0218   &       $-$0.&6082   &       $-$2.&5505   &       $-$2.&1306   \\
\hline
CI-DFS
      &       $-$0.&0586   &          0.&0289   &       $-$1.&0172   &       $-$0.&6049   &       $-$2.&5387   &       $-$2.&1241       \\
\hline
\hline
\end{tabular}
\end{table}

{
\renewcommand{\arraystretch}{0.85}
\begin{table}
\caption{Individual contributions to the ground-state ionization energy in boronlike calcium (in eV).
See text for details.}\label{table:Ca}
\bigskip
\begin{tabular}{l@{\quad}r@{\quad}r@{\quad}r@{\quad}r@{\quad}r}
\hline
\hline
\multicolumn{1}{c}{Contribution} & \multicolumn{1}{c}{CH}     &
\multicolumn{1}{c}{LDF}            & \multicolumn{1}{c}{KS}    &
\multicolumn{1}{c}{PZ1}              & \multicolumn{1}{c}{PZ3}      \\
\hline

$ E^{(0)}_{\rm Dirac}               $  &  $   -954.78362 $  &  $   -975.97881 $  &  $   -937.16983 $  &  $   -976.85080 $  &  $   -976.88741 $  \\

$ E^{(1)}_{\rm int}                 $  &  $    -21.69222 $  &  $      0.30879 $  &  $    -39.11864 $  &  $      1.74929 $  &  $      1.80159 $  \\

%

$ E^{(2)}_{\rm int}                 $  &  $      2.87596 $  &  $      1.96328 $  &  $      2.28272 $  &  $      1.30917 $  &  $      1.29162 $  \\

 $ E^{(\geqslant3)}_{\rm int,Breit} $  &  $     -0.16539 $  &  $     -0.05858 $  &  $      0.24048 $  &  $      0.02699 $  &  $      0.02886 $  \\


%

$ E^{(1)}_{\rm QED}                 $  &  $     -0.00562 $  &  $     -0.00582 $  &  $     -0.00596 $  &  $     -0.00599 $  &  $     -0.00600 $  \\

%

$ E^{(2)}_{\rm ScrQED}              $  &  $     -0.02054 $  &  $     -0.02035 $  &  $     -0.02017 $  &  $     -0.02036 $  &  $     -0.02036 $  \\

$ E^{(2)}_{\rm 2loop}               $  &  $      0.00002 $  &  $      0.00002 $  &  $      0.00002 $  &  $      0.00002 $  &  $      0.00002 $  \\


$ E_{\rm rec,Breit}                 $  &  $      0.00522 $  &  $      0.00522 $  &  $      0.00522 $  &  $      0.00522 $  &  $      0.00522 $  \\

$ E_{\rm rec,QED}                   $  &  $     -0.00001 $  &  $     -0.00001 $  &  $     -0.00001 $  &  $     -0.00001 $  &  $     -0.00001 $  \\


\hline
$ E_{\rm total}                     $  &  $   -973.78620 $  &  $   -973.78625 $  &  $   -973.78618 $  &  $   -973.78646 $  &  $   -973.78647 $  \\
\hline
\hline

\end{tabular}%
\end{table}
}

{
\renewcommand{\arraystretch}{0.85}
\begin{table}
\caption{Individual contributions to the ground-state ionization energy in boronlike xenon (in eV).
See text for details.}\label{table:Xe}
\bigskip
\begin{tabular}{l@{\quad}r@{\quad}r@{\quad}r@{\quad}r@{\quad}r}
\hline
\hline
\multicolumn{1}{c}{Contribution} & \multicolumn{1}{c}{CH}     &
\multicolumn{1}{c}{LDF}            & \multicolumn{1}{c}{KS}    &
\multicolumn{1}{c}{PZ1}              & \multicolumn{1}{c}{PZ3}      \\
\hline

$ E^{(0)}_{\rm Dirac}               $  &  $   -9189.3739 $  &  $   -9254.9556 $  &  $   -9134.5274 $  &  $   -9263.0646 $  &  $   -9264.4823 $  \\

$ E^{(1)}_{\rm int}                 $  &  $     -55.0107 $  &  $      11.2222 $  &  $    -109.5405 $  &  $      20.5477 $  &  $      22.1216 $  \\

%

$ E^{(2)}_{\rm int}                 $  &  $       7.0051 $  &  $       6.4966 $  &  $       5.6448 $  &  $       4.8744 $  &  $       4.7118 $  \\

 $ E^{(\geqslant3)}_{\rm int,Breit} $  &  $      -0.8749 $  &  $      -1.0172 $  &  $       0.1693 $  &  $      -0.6113 $  &  $      -0.6049 $  \\


%

$ E^{(1)}_{\rm QED}                 $  &  $       0.0044 $  &  $       0.0055 $  &  $       0.0013 $  &  $       0.0061 $  &  $       0.0063 $  \\

%

$ E^{(2)}_{\rm ScrQED}              $  &  $      -0.2677 $  &  $      -0.2688 $  &  $      -0.2654 $  &  $      -0.2712 $  &  $      -0.2717 $  \\

$ E^{(2)}_{\rm 2loop}               $  &  $       0.0014 $  &  $       0.0014 $  &  $       0.0014 $  &  $       0.0014 $  &  $       0.0014 $  \\


$ E_{\rm rec,Breit}                 $  &  $       0.0179 $  &  $       0.0179 $  &  $       0.0179 $  &  $       0.0179 $  &  $       0.0179 $  \\

$ E_{\rm rec,QED}                   $  &  $      -0.0004 $  &  $      -0.0004 $  &  $      -0.0004 $  &  $      -0.0004 $  &  $      -0.0004 $  \\


\hline
$ E_{\rm total}                     $  &  $   -9238.4987 $  &  $   -9238.4985 $  &  $   -9238.4988 $  &  $   -9238.5000 $  &  $   -9238.5003 $  \\
\hline
\hline

\end{tabular}%
\end{table}
}

{
\renewcommand{\arraystretch}{0.85}
\begin{table}
\caption{Individual contributions to the ground-state ionization energy in boronlike uranium (in eV).
See text for details.}\label{table:U}
\bigskip
\begin{tabular}{l@{\quad}r@{\quad}r@{\quad}r@{\quad}r@{\quad}r}
\hline
\hline
\multicolumn{1}{c}{Contribution} & \multicolumn{1}{c}{CH}     &
\multicolumn{1}{c}{LDF}            & \multicolumn{1}{c}{KS}    &
\multicolumn{1}{c}{PZ1}              & \multicolumn{1}{c}{PZ3}      \\
\hline

$ E^{(0)}_{\rm Dirac}               $  &  $  -31600.4515 $  &  $  -31721.2790 $  &  $  -31498.3316 $  &  $  -31769.4790 $  &  $  -31785.4209 $  \\

$ E^{(1)}_{\rm int}                 $  &  $     -47.6740 $  &  $      73.7308 $  &  $    -148.8099 $  &  $     124.9420 $  &  $     141.8611 $  \\

%

$ E^{(2)}_{\rm int}                 $  &  $      11.3300 $  &  $      11.1670 $  &  $       8.7078 $  &  $       7.6507 $  &  $       6.7666 $  \\

 $ E^{(\geqslant3)}_{\rm int,Breit} $  &  $      -2.1250 $  &  $      -2.5387 $  &  $      -0.4860 $  &  $      -2.0316 $  &  $      -2.1241 $  \\


%

$ E^{(1)}_{\rm QED}                 $  &  $       5.9881 $  &  $       6.0062 $  &  $       6.0244 $  &  $       6.0986 $  &  $       6.1267 $  \\

%

$ E^{(2)}_{\rm ScrQED}              $  &  $      -1.5153 $  &  $      -1.5337 $  &  $      -1.5580 $  &  $      -1.6390 $  &  $      -1.6715 $  \\

$ E^{(2)}_{\rm 2loop}               $  &  $      -0.0039 $  &  $      -0.0039 $  &  $      -0.0039 $  &  $      -0.0039 $  &  $      -0.0039 $  \\


$ E_{\rm rec,Breit}                 $  &  $       0.0540 $  &  $       0.0540 $  &  $       0.0540 $  &  $       0.0540 $  &  $       0.0540 $  \\

$ E_{\rm rec,QED}                   $  &  $      -0.0037 $  &  $      -0.0037 $  &  $      -0.0037 $  &  $      -0.0037 $  &  $      -0.0037 $  \\

$ E_{\rm nucl.pol.}                 $  &  $      -0.0042 $  &  $      -0.0042 $  &  $      -0.0042 $  &  $      -0.0042 $  &  $      -0.0042 $  \\

\hline
$ E_{\rm total}                     $  &  $  -31634.4054 $  &  $  -31634.4051 $  &  $  -31634.4111 $  &  $  -31634.4161 $  &  $  -31634.4198 $  \\
\hline
\hline

\end{tabular}%
\end{table}
}

{
\renewcommand{\arraystretch}{0.82}
\begin{table}
\caption{Ionization eneries (in eV) for boronlike ions with $Z = 16-96$.}\label{table:IP}
\begin{center}
\begin{tabular}{l@{\qquad\qquad\qquad}l@{\qquad\qquad\qquad}l}
 \hline
 \hline
Nucleus & This work & Other works \\
 \hline

  $ ^{32}_{16}$S       &    $ -564.4085(42) $    &    $ -564.407(55) ^{\rm a} $    \\
                       &                         &    $ -564.48      ^{\rm b} $    \\
                       &                         &    $ -564.46(17)  ^{\rm c} $    \\
  $ ^{35}_{17}$Cl      &    $ -656.3130(45) $    &    $ -656.30(12) ^{\rm a} $    \\
                       &                         &    $ -656.38     ^{\rm b} $    \\
                       &                         &    $ -656.37(20) ^{\rm c} $    \\
  $ ^{40}_{18}$Ar      &    $ -755.1639(46) $    &    $ -755.13(22) ^{\rm a} $    \\
                       &                         &    $ -755.22     ^{\rm b} $    \\
                       &                         &    $ -755.22(23) ^{\rm c} $    \\
                       &                         &    $ -755.227    ^{\rm d} $    \\
  $ ^{39}_{19}$K       &    $ -860.9809(48) $    &    $ -860.92(25) ^{\rm a} $    \\
                       &                         &    $ -861.03     ^{\rm b} $    \\
                       &                         &    $ -861.04(26) ^{\rm c} $    \\
  $ ^{40}_{20}$Ca      &    $ -973.7863(53) $    &    $ -973.70(33) ^{\rm a} $    \\
                       &                         &    $ -973.83     ^{\rm b} $    \\
                       &                         &    $ -973.84(29) ^{\rm c} $    \\
  $ ^{45}_{21}$Sc      &    $ -1093.6041(54) $   &    $ -1093.48(36) ^{\rm a} $    \\
                       &                         &    $ -1093.7      ^{\rm b} $    \\
                       &                         &    $ -1093.65(33) ^{\rm c} $    \\
  $ ^{48}_{22}$Ti      &    $ -1220.4583(54) $   &    $ -1220.30(46) ^{\rm a} $    \\
                       &                         &    $ -1220.5      ^{\rm b} $    \\
  $ ^{51}_{23}$V       &    $ -1354.3758(56) $   &    $ -1354.17(53) ^{\rm a} $    \\
                       &                         &    $ -1354.4      ^{\rm b} $    \\
  $ ^{52}_{24}$Cr      &    $ -1495.3846(57) $   &    $ -1495.13(66) ^{\rm a} $    \\
                       &                         &    $ -1495.4      ^{\rm b} $    \\
  $ ^{55}_{25}$Mn      &    $ -1643.5149(60) $   &    $ -1643.21(72) ^{\rm a} $    \\
                       &                         &    $ -1643.5      ^{\rm b} $    \\
  $ ^{56}_{26}$Fe      &    $ -1798.7978(63) $   &    $ -1798.43(78) ^{\rm a} $    \\
                       &                         &    $ -1798.8      ^{\rm b} $    \\
  $ ^{59}_{27}$Co      &    $ -1961.2669(65) $   &    $ -1960.83(84) ^{\rm a} $    \\
                       &                         &    $ -1961.3      ^{\rm b} $    \\
  $ ^{58}_{28}$Ni      &    $ -2130.9563(68) $   &    $ -2130.46(92) ^{\rm a} $    \\
                       &                         &    $ -2131.0      ^{\rm b} $    \\
  $ ^{63}_{29}$Cu      &    $ -2307.9038(72) $   &    $ -2307.3(1.0) ^{\rm a} $    \\
                       &                         &    $ -2307.9      ^{\rm b} $    \\

\end{tabular}
\end{center}
\end{table}
}
\addtocounter{table}{-1}
{
\renewcommand{\arraystretch}{0.82}
\begin{table}
\caption{(Continued.)}
\begin{center}
\begin{tabular}{l@{\qquad\qquad\qquad}l@{\qquad\qquad\qquad}l}
 \hline
 \hline
Nucleus & This work & Other works \\
 \hline

  $ ^{64}_{30}$Zn      &    $ -2492.1465(74) $    &    $ -2491.5(1.1) ^{\rm a} $    \\
                       &                          &    $ -2492.1      ^{\rm b}  $    \\
  $ ^{69}_{31}$Ga      &    $ -2683.7257(77) $    &    $ -2683.0(1.4) ^{\rm a} $    \\
                       &                          &    $ -2683.7      ^{\rm b} $    \\
  $ ^{74}_{32}$Ge      &    $ -2882.6828(81) $    &    $ -2881.9(1.6) ^{\rm a} $    \\
                       &                          &    $ -2882.7      ^{\rm b} $    \\
  $ ^{75}_{33}$As      &    $ -3089.0615(84) $    &    $ -3088.1(1.9) ^{\rm a} $    \\
                       &                          &    $ -3089.0      ^{\rm b} $    \\
  $ ^{80}_{34}$Se      &    $ -3302.9084(88) $    &    $ -3301.8(2.1) ^{\rm a} $    \\
                       &                          &    $ -3302.9      ^{\rm b} $    \\
  $ ^{79}_{35}$Br      &    $ -3524.2705(91) $    &    $ -3523.1(2.2) ^{\rm a} $    \\
                       &                          &    $ -3524.2      ^{\rm b} $    \\
  $ ^{84}_{36}$Kr      &    $ -3753.1991(94) $    &    $ -3752.0(2.5) ^{\rm a} $    \\
                       &                          &    $ -3753.1      ^{\rm b} $    \\
  $ ^{85}_{37}$Rb      &    $ -3989.745(10) $     &    $ -3988.4(2.7) ^{\rm a} $    \\
                       &                          &    $ -3989.7      ^{\rm b} $    \\
  $ ^{88}_{38}$Sr      &    $ -4233.963(10) $     &    $ -4232.4(3.0) ^{\rm a} $    \\
                       &                          &    $ -4233.9      ^{\rm b} $    \\
  $ ^{89}_{39}$Y       &    $ -4485.910(11) $     &    $ -4484.4(3.2) ^{\rm a} $    \\
                       &                          &    $ -4485.8      ^{\rm b} $    \\
  $ ^{90}_{40}$Zr      &    $ -4745.644(11) $     &    $ -4743.9(3.5) ^{\rm a} $    \\
                       &                          &    $ -4745.5      ^{\rm b} $    \\
  $ ^{93}_{41}$Nb      &    $ -5013.226(12) $     &    $ -5011.4(3.7) ^{\rm a} $    \\
                       &                          &    $ -5013.1      ^{\rm b} $    \\
  $ ^{98}_{42}$Mo      &    $ -5288.721(12) $     &    $ -5286.8(4.0) ^{\rm a} $    \\
                       &                          &    $ -5288.6      ^{\rm b} $    \\
  $ ^{97}_{43}$Tc      &    $ -5572.193(13) $     &    $ -5570.1(4.2) ^{\rm a} $    \\
                       &                          &    $ -5572.1      ^{\rm b} $    \\
  $^{102}_{44}$Ru      &    $ -5863.711(13) $     &    $ -5861.5(4.5) ^{\rm a} $    \\
                       &                          &    $ -5863.6      ^{\rm b} $    \\
  $^{103}_{45}$Rh      &    $ -6163.347(14) $     &    $ -6161.0(4.7) ^{\rm a} $    \\
                       &                          &    $ -6163.2      ^{\rm b} $    \\
  $^{106}_{46}$Pd      &    $ -6471.174(14) $     &    $ -6468.6(5.0) ^{\rm a} $    \\
                       &                          &    $ -6471.0      ^{\rm b} $    \\
  $^{107}_{47}$Ag      &    $ -6787.269(15) $     &    $ -6784.5(5.2) ^{\rm a} $    \\
                       &                          &    $ -6787.0      ^{\rm b} $    \\

\end{tabular}
\end{center}
\end{table}
}
\addtocounter{table}{-1}
{
\renewcommand{\arraystretch}{0.82}
\begin{table}
\caption{(Continued.)}
\begin{center}
\begin{tabular}{l@{\qquad\qquad\qquad}l@{\qquad\qquad\qquad}l}
 \hline
 \hline
Nucleus & This work & Other works \\
 \hline

  $^{114}_{48}$Cd      &    $ -7111.712(15) $    &    $ -7108.9(5.3) ^{\rm a} $    \\
                       &                         &    $ -7111.5      ^{\rm b} $    \\
  $^{115}_{49}$In      &    $ -7444.584(16) $    &    $ -7441.7(5.6) ^{\rm a} $    \\
                       &                         &    $ -7444.4      ^{\rm b} $    \\
  $^{120}_{50}$Sn      &    $ -7785.972(17) $    &    $ -7785.8      ^{\rm b} $    \\
                       &                         &    $ -7785.271    ^{\rm d} $    \\
                       &                         &    $ -7790(4)     ^{\rm e} $    \\
  $^{121}_{51}$Sb      &    $ -8135.963(17) $    &    $ -8135.7      ^{\rm b} $    \\
                       &                         &    $ -8140(4)     ^{\rm e} $    \\
  $^{130}_{52}$Te      &    $ -8494.651(18) $    &    $ -8494.4      ^{\rm b} $    \\
                       &                         &    $ -8499(4)     ^{\rm e} $    \\
  $^{127}_{53}$I       &    $ -8862.130(19) $    &    $ -8861.9      ^{\rm b} $    \\
                       &                         &    $ -8867(4)     ^{\rm e} $    \\
  $^{132}_{54}$Xe      &    $ -9238.499(20) $    &    $ -9238.2      ^{\rm b} $    \\
                       &                         &    $ -9243(4)     ^{\rm e} $    \\
  $^{133}_{55}$Cs      &    $ -9623.861(20) $    &    $ -9623.6      ^{\rm b} $    \\
                       &                         &    $ -9629(4)     ^{\rm e} $    \\
  $^{138}_{56}$Ba      &    $ -10018.323(21) $   &    $ -10018       ^{\rm b} $    \\
                       &                         &    $ -10023(4)    ^{\rm e} $    \\
  $^{139}_{57}$La      &    $ -10421.995(22) $   &    $ -10422      ^{\rm b} $    \\
                       &                         &    $ -10426(4)   ^{\rm e} $    \\
  $^{140}_{58}$Ce      &    $ -10834.990(23) $   &    $ -10835      ^{\rm b} $    \\
                       &                         &    $ -10840(5)   ^{\rm e} $    \\
  $^{141}_{59}$Pr      &    $ -11257.428(24) $   &    $ -11257      ^{\rm b} $    \\
                       &                         &    $ -11262(5)   ^{\rm e} $    \\
  $^{142}_{60}$Nd      &    $ -11689.432(26) $   &    $ -11689      ^{\rm b} $    \\
                       &                         &    $ -11694(5)   ^{\rm e} $    \\
  $^{145}_{61}$Pm      &    $ -12131.130(27) $   &    $ -12136(5)   ^{\rm e} $    \\
  $^{152}_{62}$Sm      &    $ -12582.654(28) $   &    $ -12588(6)   ^{\rm e} $    \\
  $^{153}_{63}$Eu      &    $ -13044.142(30) $   &    $ -13050(6)   ^{\rm e} $    \\
  $^{158}_{64}$Gd      &    $ -13515.737(31) $   &    $ -13521(7)   ^{\rm e} $    \\
  $^{159}_{65}$Tb      &    $ -13997.589(33) $   &    $ -14003(10)  ^{\rm e} $    \\
  $^{164}_{66}$Dy      &    $ -14489.846(34) $   &    $ -14495(12)  ^{\rm e} $    \\
  $^{165}_{67}$Ho      &    $ -14992.676(36) $   &    $ -14998(15) ^{\rm e}  $    \\
  $^{166}_{68}$Er      &    $ -15506.240(38) $   &    $ -15511(19)  ^{\rm e} $    \\

\end{tabular}
\end{center}
\end{table}
}
\addtocounter{table}{-1}
{
\renewcommand{\arraystretch}{0.82}
\begin{table}
\caption{(Continued.)}
\begin{center}
\begin{tabular}{l@{\qquad\qquad\qquad}l@{\qquad\qquad\qquad}l}
 \hline
 \hline
Nucleus & This work & Other works \\
 \hline

  $^{169}_{69}$Tm      &    $ -16030.716(40) $    &    $ -16036(24)  ^{\rm e} $    \\
  $^{174}_{70}$Yb      &    $ -16566.277(42) $    &    $ -16572(30)  ^{\rm e} $    \\
  $^{175}_{71}$Lu      &    $ -17113.113(44) $    &    $ -17118(40)  ^{\rm e} $    \\
  $^{180}_{72}$Hf      &    $ -17671.425(46) $    &    $ -17677(40)  ^{\rm e} $    \\
  $^{181}_{73}$Ta      &    $ -18241.409(48) $    &    $ -18247(50)  ^{\rm e} $    \\
  $^{184}_{74}$W       &    $ -18823.276(51) $    &    $ -18829(50)  ^{\rm e} $    \\
  $^{187}_{75}$Re      &    $ -19417.250(53) $    &    $ -19423(60)  ^{\rm e} $    \\
  $^{192}_{76}$Os      &    $ -20023.553(56) $    &    $ -20030(60)  ^{\rm e} $    \\
  $^{193}_{77}$Ir      &    $ -20642.435(60) $    &    $ -20648(70)  ^{\rm e} $    \\
  $^{194}_{78}$Pt      &    $ -21274.135(62) $    &    $ -21280(90)  ^{\rm e} $    \\
  $^{197}_{79}$Au      &    $ -21918.920(65) $    &    $ -21925(90)  ^{\rm e} $    \\
  $^{202}_{80}$Hg      &    $ -22577.057(68) $    &    $ -22583(100) ^{\rm e} $    \\
  $^{205}_{81}$Tl      &    $ -23248.837(72) $    &    $ -23255(110) ^{\rm e} $    \\
  $^{208}_{82}$Pb      &    $ -23934.548(76) $    &    $ -23941(120) ^{\rm e} $    \\
  $^{209}_{83}$Bi      &    $ -24634.509(80) $    &    $ -24641(120) ^{\rm e} $    \\
  $^{210}_{84}$Po      &    $ -25349.032(84) $    &    $ -25355(140) ^{\rm e} $    \\
  $^{215}_{85}$At      &    $ -26078.481(92) $    &    $ -26085(150) ^{\rm e} $    \\
  $^{220}_{86}$Rn      &    $ -26823.168(94) $    &    $ -26829(160) ^{\rm e} $    \\
  $^{223}_{87}$Fr      &    $ -27583.52(10) $     &    $ -27590(170) ^{\rm e} $    \\
  $^{226}_{88}$Ra      &    $ -28359.91(11) $     &    $ -28366(190) ^{\rm e} $    \\
  $^{227}_{89}$Ac      &    $ -29152.80(12) $     &    $ -29159(200) ^{\rm e} $    \\
  $^{232}_{90}$Th      &    $ -29962.47(12) $     &    $ -29968(220) ^{\rm e} $    \\
  $^{231}_{91}$Pa      &    $ -30789.62(13) $     &    $ -30796(240) ^{\rm e} $    \\
  $^{238}_{92}$U       &    $ -31634.41(13) $     &    $ -31641(250) ^{\rm e} $    \\
  $^{237}_{93}$Np      &    $ -32497.76(15) $     &    $ -32504(300) ^{\rm e} $    \\
  $^{240}_{94}$Pu      &    $ -33379.74(15) $     &    $ -33386(300) ^{\rm e} $    \\
  $^{243}_{95}$Am      &    $ -34281.24(16) $     &    $ -34287(300) ^{\rm e} $    \\
  $^{244}_{96}$Cm      &    $ -35202.90(17) $     &    $ -35209(300) ^{\rm e} $    \\

\hline
\end{tabular}
\end{center}
\bigskip
\raggedright

$^{\rm a}$ E.~Bi\'emont \textit{et al.} \cite{Biemont:1999:117}.

$^{\rm b}$ M.~F.~Gu \cite{Gu:2005:267}.

$^{\rm c}$ N.~N.~Dutta and S.~Majumder \cite{Dutta:2012:032512} (in this work the QED corrections are not taken into account).

$^{\rm d}$ E.~Eliav \textit{et al.} \cite{Eliav:1994:82} (in this work the QED corrections are not taken into account).

$^{\rm e}$ G.~C.~Rodrigues \textit{et al.} \cite{Rodrigues:2004:117} with the uncertainty
prescribed by NIST \cite{NIST:2015}.

\end{table}
}

In the present section we discuss our results for the ground-state ionization energies of boronlike ions.
The interelectronic-interaction contributions of the third and higher orders
have been evaluated in the Breit approximation by means of two independent approaches, namely, with the use of
the CI-DFS method and the recursive formulation of the perturbation theory.
In Table~\ref{table:PT} we show the results of our calculations within both methods for
boronlike calcium, xenon, and uranium which were obtained employing the LDF and PZ3 screening potentials
as the zeroth-order approximations. One can see that in all the cases the results are in a good agreement
with each other.
In the final compilation we have used the data evaluated within the CI-DFS approach.
The deviation of the CI-DFS results from the ones obtained by the
perturbation theory was used to estimate the uncertainty associated with this correction.

The individual contributions to the ground-state ionization energies of boronlike calcium, xenon, and uranium
calculated for the five different effective potentials are given in Tables~\ref{table:Ca}-\ref{table:U}, respectively.
The zeroth-order approximation for the ionization energies obtained from the one-electron Dirac equation~(\ref{DirEq}) is
presented for each ion in the first line. For uranium we added the nuclear deformation correction in accordance
with~\cite{Kozhedub:2008:032501}. In the second and third rows we give the interelectronic-interaction corrections
which correspond to the first- and second-order diagrams in Fig.~\ref{fig:int}, respectively.
These contributions were evaluated within the rigorous QED approach according to the formulas presented in
Appendix~\ref{app:A}. The fourth line contains the electron-correlation corrections of the third and higher orders
in the Breit approximation obtained from the CI-DFS calculations.
In the next two lines we give the contributions of the one-electron and many-electron one-loop QED
diagrams depicted in Fig.~\ref{fig:qed}.
The two-loop one-electron QED corrections are presented in the seventh row.
The next two lines display the Breit and QED parts of the nuclear recoil correction.
In Table~\ref{table:U}, for uranium we have included the contribution due to the nuclear polarization
effect $E_{\rm nucl.pol.}$. Finally, the total values of the ionization energies are
presented in the last line.
From Tables~\ref{table:Ca}-\ref{table:U} one can see that the final results for different screening
potentials deviate from each other much less than the results with the many-electron QED effects neglected.
Therefore, the present calculations provide much better accuracy than that of all previous studies
of the ground-state ionization energies in B-like ions.
For all other ions in the range $16 \leqslant Z \leqslant 96$
we have performed the evaluation of the ground-state ionization energies
using the LDF and PZ3 screening potentials only. These two types of the effective potential
have been chosen in order to control the accuracy of the calculations along the whole boron
isoelectronic sequence.

In Table~\ref{table:IP} we present the ground-state ionization energies for all boronlike ions
in the range $16 \leqslant Z \leqslant 96$. For calcium, xenon, and uranium the final values
were obtained by averaging the results of the calculations with the five effective potentials.
For all other ions the averages of the total values evaluated with the use of the
LDF and PZ3 screening potentials are presented.
The uncertainties of our theoretical calculations are given in the parentheses.
We obtained them by summing quadratically the uncertainty due to the nuclear size effect,
the uncertainty of the $E^{(\geqslant3)}_{\rm int,Breit}$ term and the uncertainty
associated with the uncalculated higher-order QED contributions.
Since the wave function of the $2p_{1/2}$ state is small in the nuclear area,
the uncertainty of the nuclear size correction contributes significantly only for high-$Z$ ions.
For uranium this uncertainty was estimated according to Ref.~\cite{Kozhedub:2008:032501}.
For all other ions in order to calculate the uncertainty of the nuclear size correction
we added quadratically two uncertainties.
The first one was obtained by varying
the root-mean-square nuclear radius within its error bar. The second uncertainty
was evaluated by changing the nuclear charge distribution model from the Fermi model
to the homogeneously charged sphere model.

In Table~\ref{table:IP} our results for the ionization energies of boronlike ions
are compared with theoretical predictions made by other groups.
One can see that our results, as a rule, are in a good agreement with the results
of the previous relativistic calculations.
However, in contrast to all previous evaluations of the ground-state energies of B-like ions,
we have accounted for the many-electron QED effects rigorously and did not use any one-electron
approximations or semiempirical approaches.
As a result, in our work the theoretical accuracy for the ionization energies of the $2p_{1/2}$
electron in boronlike ions has been drastically improved.


\section{Summary \label{sec:3}}

To summarize, the high-precision QED calculations of the ground-state ionization energies
for all boronlike ions in the range $16 \leqslant Z \leqslant 96$ have been performed.
The contributions of all Feynman diagrams up to the second order of the perturbation theory are taken into account.
The many-electron QED effects are rigorously evaluated in the framework of the
extended Furry picture without an expansion
in powers of the interaction with the effective potential.
The third- and higher-order correlation effects are accounted for within the Breit approximation.
The contributions of the nuclear recoil effect are also evaluated.
As the result, the most precise theoretical predictions for the ionization energies of boronlike ions
have been obtained. The achieved accuracy gives an opportunity to probe QED corrections
in the ionization energies of B-like ions. 


\section*{Acknowledgements}
This work was supported by RFBR (Grants No.~16-02-00334, No.~15-03-07644, and No.~17-02-00216), by SPSU
(Grants No.~11.38.237.2015, No.~11.42.665.2017, No.~11.42.688.2017, No.~11.42.668.2017, and No.~11.42.666.2017),
and by SPSU-DFG (Grants No.~11.65.41.2017 and No.~STO~346/5-1).
A.V.M. acknowledges the support from the German Academic Exchange Service (DAAD),
from TU Dresden (DAAD-Programm Ostpartnerschaften).
The work was carried out with the financial support of the FAIR-Russia
Research Center.


\appendix


\section{Interelectronic-interaction corrections within QED \label{app:A}}

The calculation formulas derived within the TTGF method~\cite{TTGF} for the contributions of the first- and second-order
interelectronic-interaction diagrams shown in Fig.~\ref{fig:int}
are collected in the present Appendix.

The expressions for the contributions of the first-order diagrams~(a) and (e)
in Fig.~\ref{fig:int} are well-known
\begin{gather}
\Delta E_{\rm \ref{fig:int}(a)} = \sum_{c=1s,2s} \sum_{\mu_c} \sum_P
(-1)^P I_{PcPvcv}(\veps_{Pc}-\veps_c),
		\label{eq:int:a} \\[1mm]
\Delta E_{\rm \ref{fig:int}(e)} = V_{vv},
		\label{eq:int:e}
\end{gather}
where $c$ corresponds to the states of the core electrons, $v$ denotes the $2p_{1/2}$ valence electron state,
$\mu$ stands for the angular momentum projection of the electron state,
$I_{abcd}(\omega) = \langle ab| I(\omega)|cd\rangle$, $I(\omega)=e^2 \alpha^\mu \alpha^\nu D_{\mu\nu}(\omega)$,
$D_{\mu\nu}(\omega)$ is the photon propagator,
$V_{ab}=\langle a | \delta V | b \rangle$, $\delta V =-V_{\rm{scr}}$,
$P$ denotes the permutation operator, and $(-1)^P$ is the sign of the permutation.

The derivation of the formulas for the second-order corrections arising from the
two-photon exchange diagrams~(b)-(d) in Fig.~\ref{fig:int}
has been discussed in details in Refs.~\cite{Shabaev:1994:4489,Yerokhin:2001:032109,Artemyev:2003:062506}.
For convenience, the final expressions needed for calculation of the ionization energies of boronlike ions
are given below.
The contributions of these diagrams are divided naturally into two parts: reducible and irreducible.
The reducible part includes
the terms in which an intermediate-state energy in the diagram coincides with the energy of the initial (final) state.
The irreducible part corresponds to the remainder.
First, we will consider the contributions of the three-electron diagram~(d).
The irreducible (``irr'') and reducible (``red'') parts of this correction for
a three-electron configuration $(1,2,3)$ read as follows \cite{Yerokhin:2001:032109},
\begin{eqnarray}
\Delta E_{\rm irr}^{\rm 3el}(1,2,3)
  &=& \sum_{PQ} (-1)^{P+Q} \sum_{\veps_n\neq\veps_{Q1}+\veps_{Q2}-\veps_{P1}}  
  \,\frac{I_{P2P3nQ3}(\Delta_{P3Q3})I_{P1nQ1Q2}(\Delta_{Q1P1})}
                                       {\veps_{Q1}+\veps_{Q2}-\veps_{P1}-\veps_n},
		\label{eq:2ph_3el_irr} \\[1mm]
\Delta E_{\rm red}^{\rm 3el}(1,2,3) &=& \frac{1}{2} \sum_{PQ} (-1)^{P+Q}
     \sum_{\veps_n=\veps_{Q1}+\veps_{Q2}-\veps_{P1}}
              \nonumber \\[1mm]
    &&   \left[ I'_{P2P3nQ3}(\Delta_{P3Q3})I_{P1nQ1Q2}(\Delta_{Q1P1})
              + I_{P2P3nQ3}(\Delta_{P3Q3})I'_{P1nQ1Q2}(\Delta_{Q1P1}) \right],
		\label{eq:2ph_3el_red}
\end{eqnarray}
where $I'_{abcd}(\omega)=\langle ab| \tfrac{\partial}{\partial \omega}I(\omega)|cd\rangle$,
$\Delta_{ab} = \varepsilon_a - \varepsilon_b$, $P$ and $Q$ are the permutation operators.
Therefore, for the total three-electron contribution to the ground-state ionization energy of boronlike ions we obtain
\begin{equation}
\Delta E_{\rm \ref{fig:int}(d)} =
     \sum_{c=1s,2s} \left[ \Delta E_{\rm irr}^{\rm 3el}(c,\bar{c},v) + \Delta E_{\rm red}^{\rm 3el}(c,\bar{c},v) \right]
               + \sum_{\mu_a,\mu_b} \left[ \Delta E_{\rm irr}^{\rm 3el}(a,b,v)       + \Delta E_{\rm red}^{\rm 3el}(a,b,v)       \right],
\label{eq:2ph_3el_tot_B_IP}
\end{equation}
where $\bar{c}$ denotes the electron state with $\veps_{\bar{c}}=\veps_c$ and $\mu_{\bar{c}}=-\mu_c$,
$a=1s$, and $b=2s$. The first sum in (\ref{eq:2ph_3el_tot_B_IP}) describes the interaction of the valence electron
with two core electrons $c$ and $\bar{c}$ belonging to the same electron shell,
the second sum corresponds the case when the core electrons $a$ and $b$
are from the different shells.

The contributions of the two-electron diagrams~(b) and (c)
are referred to as the ladder (``lad'') and crossed (``cr'') terms respectively.
The sum of the infrared-finite part of the crossed term and the irreducible part of the ladder term can be expressed as
\begin{eqnarray}
\Delta E^{\rm 2el}_{1} &=& \sum_{c=1s,2s}\,\, \sideset{}{'}\sum_{n_1,n_2}{} \,\, \frac{i}{2\pi} \int_{-\infty}^{+\infty} \! d\omega \,
     \left \{ \frac{F^{\rm lad}_{\rm dir}(\omega,n_1n_2)}
           {(\veps_c-\omega-u\veps_{n_1})(\veps_v+\omega-u\veps_{n_2})} \right. \nonumber \\[1mm]
   && + \, \frac{F^{\rm lad}_{\rm exc}(\omega,n_1n_2)}
           {(\veps_v-\omega-u\veps_{n_1})(\veps_c+\omega-u\veps_{n_2})} \nonumber \\[1mm]
   && + \, \frac{F^{\rm cr}_{\rm dir}(\omega,n_1n_2)}
           {(\veps_c-\omega-u\veps_{n_1})(\veps_v-\omega-u\veps_{n_2})} \nonumber \\[1mm]
   && \left. + \, \frac{F^{\rm cr}_{\rm exc}(\omega,n_1n_2)}
           {(\veps_v-\omega-u\veps_{n_1})(\veps_v-\omega-u\veps_{n_2})} \right \},
\label{eq:2ph_irr}
\end{eqnarray}
where
\begin{eqnarray}
 F^{\rm lad}_{\rm dir}(\omega,n_1n_2) &=& \sum_{\mu_c} \sum_{\mu_{n_1}\mu_{n_2}}
       I_{cvn_1n_2}(\omega) I_{n_1n_2cv}(\omega), \nonumber \\[1mm]
-F^{\rm lad}_{\rm exc}(\omega,n_1n_2) &=& \sum_{\mu_c} \sum_{\mu_{n_1}\mu_{n_2}}
       I_{vcn_1n_2}(\omega) I_{n_1n_2cv}(\omega-\Delta_{vc}), \nonumber \\[1mm]
 F^{\rm  cr}_{\rm dir}(\omega,n_1n_2) &=& \sum_{\mu_c} \sum_{\mu_{n_1}\mu_{n_2}}
       I_{cn_2n_1v}(\omega) I_{n_1vcn_2}(\omega), \nonumber \\[1mm]
-F^{\rm  cr}_{\rm exc}(\omega,n_1n_2) &=& \sum_{\mu_c} \sum_{\mu_{n_1}\mu_{n_2}}
       I_{vn_2n_1v}(\omega) I_{n_1ccn_2}(\omega-\Delta_{vc}),
\label{eq:2ph_irr_F}
\end{eqnarray}
$u=(1-i0)$ provides the proper treatment of the poles in the electron propagator,
and the prime on the sum in (\ref{eq:2ph_irr}) indicates that several terms are excluded
from the summation. First of all, the states contributing to the reducible part of the ladder correction are omitted,
i.e., the terms in which the intermediate two-electron energy $\veps_{n_1}+\veps_{n_2}$ coincides with the initial
two-electron energy $\veps_c+\veps_v$. In addition, the infrared-divergent terms of the crossed contribution
are excluded, namely, the terms with $(\veps_{n_1}\veps_{n_2})=(\veps_c\veps_v)$ in the direct crossed part
and the terms with $(\veps_{n_1}\veps_{n_2})=(\veps_c\veps_c)$ and $(\veps_{n_1}\veps_{n_2})=(\veps_v\veps_v)$
in the exchange crossed part (see Ref.~\cite{Shabaev:1994:4489} for details).
The singular terms should be considered together with the reducible
contribution of the ladder diagram. The sum of these contributions reads as follows
\begin{eqnarray}
\Delta E^{\rm 2el}_{2} &=& \sum_{c=1s,2s}\, \frac{i}{4\pi} \int_{-\infty}^{+\infty} \! d\omega \,
\frac{1}{(\omega+i0)^2} \left[ 2 F^{\rm cr}_{\rm exc}(-\omega+\Delta_{vc},cc) + 2 F^{\rm cr}_{\rm exc}(-\omega,vv) \right. \nonumber \\[1mm]
   && - \, F^{\rm lad}_{\rm exc}( \omega+\Delta_{vc},cv) - F^{\rm lad}_{\rm exc}(-\omega+\Delta_{vc},cv)
         - F^{\rm lad}_{\rm dir}( \omega-\Delta_{vc},vc) \nonumber \\[1mm]
   && \left. - \, F^{\rm lad}_{\rm dir}(-\omega-\Delta_{vc},vc) - F^{\rm lad}_{\rm exc}( \omega,vc)
             - F^{\rm lad}_{\rm exc}(-\omega,vc) \right].
\label{eq:2ph_red}
\end{eqnarray}
Therefore, for the sum of the corrections corresponding to the two-electron diagrams~(b) and (c) we obtain
\begin{equation}
\Delta E_{\rm \ref{fig:int}(b)} + \Delta E_{\rm \ref{fig:int}(c)} =
\Delta E^{\rm 2el}_{1} + \Delta E^{\rm 2el}_{2}.
\label{eq:2ph_2el_tot_B_IP}
\end{equation}

Finally, we have to consider the second-order counterterm diagrams~(f) and (g) in Fig.~\ref{fig:int}:
\begin{eqnarray}
\Delta E_{\rm \ref{fig:int}(f)} &=& \sum_{n \neq v} \frac{\left| V_{vn} \right|^2}{\veps_v-\veps_n},
		\label{contr_f} \\[1mm]
\Delta E_{\rm \ref{fig:int}(g)} &=& \sum_{c=1s,2s} \sum_{\mu_c} \Bigg\{ 2 \sum_P (-1)^P
                  \Bigg[ \sum_{n \neq v} \frac{I_{PcPvcn}(\veps_{Pc}-\veps_c)V_{nv}}{\veps_v-\veps_n}
                       + \sum_{n \neq c} \frac{I_{PcPvnv}(\veps_{Pv}-\veps_v)V_{nc}}{\veps_c-\veps_n} \Bigg] \nonumber \\[1mm]
  &&                   + (V_{cc}-V_{vv})I'_{cvvc}(\veps_v-\veps_c) \Bigg\}.
		\label{contr_g}
\end{eqnarray}


\section{QED corrections \label{app:B}}

The formal expressions for the first-order QED diagrams and for the second-order screening QED diagrams
shown in Fig.~\ref{fig:qed} are given in the present Appendix.

For the description of the self-energy contributions it is convenient to introduce
the SE operator $\Sigma(E)$ as follows
\begin{eqnarray}
\langle a | \Sigma(E) | b \rangle &=&
   2 i \alpha \! \int_{-\infty}^\infty \! d\omega \! \int \! d\br_1d\br_2 \,\,
   \psi_a^\dagger(\br_1) \alpha^\mu
            G(E-\omega,\br_1,\br_2) \alpha^\nu \psi_b(\br_2) D_{\mu\nu}(\omega,\br_1-\br_2) \nonumber \\[1mm]
  &=&   \frac{i}{2\pi} \int_{-\infty}^\infty \! d\omega \,
         \sum_n \frac{\langle an| e^2 \alpha^\mu\alpha^\nu D_{\mu\nu}(\omega) |nb\rangle }
           { E - \omega -\veps_n(1-i0)},
		\label{eq:SE_sigma}
\end{eqnarray}
where $D_{\rho\sigma}(\omega)$ is the photon propagator.
With this definition the first-order self-energy correction corresponding to the diagram~(a)
in Fig.~\ref{fig:qed} can be written in the form
\begin{equation}
\Delta E_{\rm \ref{fig:qed}(a)} = \langle v | \Sigma(\veps_v) | v \rangle.
\label{eq:SE_shift}
\end{equation}
The SE contribution~(\ref{eq:SE_shift}) suffers from the ultraviolet divergences.
It has to be regularized together with the mass counterterm (the diagrams for the mass counterterms
are omitted in Fig.~\ref{fig:qed}). The prescriptions how to deal with the divergences in the SE diagram are discussed
in details in Refs.~\cite{Mohr:1974:26:note,Snyderman:1991:43,Yerokhin:1999:800}.
In the present work we follow the renormalization procedure described there.

The derivation within the TTGF method of the corrections which correspond
to the screening self-energy diagrams~(b) and (d) and the renormalization procedure
for these contributions were discussed in details in Ref.~\cite{Yerokhin:1999:3522}.
The ``vertex'' diagram~(b) gives rise to the energy shift
\begin{eqnarray}
\label{eq:scrse_ver_dE}
\Delta E_{\rm \ref{fig:qed}(b)} &=&
      \sum_{c=1s,2s}\sum_{\mu_c}
      \sum_{P} (-1)^P \sum_{n_1n_2} \frac{i}{2\pi}
      \int \!  d\omega \,  \left[
        \frac{\langle n_1\,Pc |I(\Delta)|n_2\,c \rangle \langle Pv\,n_2 |I(\omega)| n_1\,v \rangle}
       {(\veps_{Pv}-\omega-u\veps_{n_1})(\veps_v-\omega-u\veps_{n_2})} \right. \nonumber \\[1mm]
  && + \left. \frac{\langle Pv\,n_1 |I(\Delta)|v\,n_2 \rangle \langle Pc\,n_2 |I(\omega)| n_1\,c \rangle}
       {(\veps_{Pc}-\omega-u\veps_{n_1})(\veps_c-\omega-u\veps_{n_2})} \right],
\end{eqnarray}
where $\Delta=\veps_{Pv}-\veps_v$.
The contribution of the diagram~(d) in Fig.~\ref{fig:qed} is represented by the expression
\begin{eqnarray}
\label{eq:scrse_irr_red}
\Delta E_{\rm \ref{fig:qed}(d)} &=& \sum_{c=1s,2s} \sum_{\mu_c} \Big[
       2 \Big\{ \langle v | \Sigma(\veps_v) | \xi_v \rangle + \langle c | \Sigma(\veps_c) | \xi_c \rangle \Big\}  \nonumber \\[1mm]
  &&    +  \langle cv | I'(\veps_v-\veps_c) | vc \rangle
   \Big\{ \langle c | \Sigma(\veps_c) | c \rangle - \langle v | \Sigma(\veps_v) | v \rangle \Big\}  \nonumber \\[1mm]
  &&    +  \sum_{P}(-1)^P \langle PvPc | I(\Delta) | vc \rangle
   \Big\{ \langle v |\Sigma'(\veps_v)| v \rangle + \langle c | \Sigma'(\veps_c) | c \rangle \Big\}  \Big],
\end{eqnarray}
where $\Sigma'(\veps_a) = d\Sigma(E) / dE |_{E=\veps_a}$ and we have introduced the wave functions
\begin{eqnarray}
| \xi_v \rangle &=& \sum_{\veps_n\neq\veps_v} \frac{|n\rangle}{\veps_v-\veps_n}
        \sum_P (-1)^P \langle nc | I(\Delta) | PvPc \rangle,
		\label{eq:scrse_irr_dE_func_v} \\[1mm]
| \xi_c \rangle &=& \sum_{\veps_n\neq\veps_c} \frac{|n\rangle}{\veps_c-\veps_n}
        \sum_P (-1)^P \langle vn | I(\Delta) | PvPc \rangle.
        \label{eq:scrse_irr_dE_func_c}
\end{eqnarray}

The derivation of the formulas for the counterterm screening SE diagrams~(c) and (e) in Fig.~\ref{fig:qed}
can be performed in a similar way. These diagrams lead to the contributions
\begin{eqnarray}
\Delta E_{\rm \ref{fig:qed}(c)} &=& \frac{i}{2\pi} \int_{-\infty}^\infty \! d\omega \, \sum_{n_1n_2}
   \frac{\langle n_1 | \delta V | n_2 \rangle  \langle v n_2 | I(\omega) | n_1 v \rangle}
   {(\veps_v-\omega-u\veps_{n_1})(\veps_v-\omega-u\veps_{n_2})},
		\label{eq:SE_CT_ver} \\[1mm]
\Delta E_{\rm \ref{fig:qed}(e)} &=& 2 \langle v | \Sigma(\veps_v) | \chi_v \rangle
   +  \langle v | \delta V | v \rangle  \langle v | \Sigma'(\veps_v) | v \rangle,
		\label{eq:SE_CT_irr_red}
\end{eqnarray}
where
\begin{eqnarray}
| \chi_v \rangle &=& \sum_{\veps_n\neq\veps_v} \frac{|n\rangle}{\veps_v-\veps_n}
                  \langle n | \delta V| v \rangle     \label{eq:scrse_irr_dE_func_w}
\end{eqnarray}
is the correction to the wave function of the $2p_{1/2}$ valence electron
owing to the interaction with the counterterm potential~$\delta V = -V_{\rm scr}$.

The energy contribution of the first-order vacuum-polarization diagram~(f) in Fig.~\ref{fig:qed}
can be written in the form
\begin{equation}
\Delta E_{\rm \ref{fig:qed}(f)} =  \langle v | U_{\rm VP} | v \rangle,
		\label{eq:VP_1el}
\end{equation}
where the VP potential $U_{\rm VP}$ is introduced as follows
\begin{equation}
U_{\rm VP}(\bx) =  \frac{\alpha}{2\pi i} \int \! d\br \, \frac{1}{|\bx-\br|}
          \int_{-\infty}^\infty \! d\omega \,  {\rm Tr} \left[  G(\omega, \br,\br) \right].
		\label{eq:VPpot}
\end{equation}

The expressions for the second-order VP diagrams~(g) and (i) can be obtained easily
in the framework of the TTGF method, see Refs.~\cite{Artemyev:1997:3529,Artemyev:1999:45,Artemyev:2005:062104}
for details,
\begin{eqnarray}
\Delta E_{\rm \ref{fig:qed}(g)} &=& \sum_{c=1s,2s} \sum_{\mu_c}
      \sum_{P} (-1)^P  \langle Pv Pc | I_{\rm VP}(\Delta) | vc \rangle, \label{eq:ScrVP_b} \\[1mm]
\Delta E_{\rm \ref{fig:qed}(i)} &=& \sum_{c=1s,2s} \sum_{\mu_c} \Big[
      2 \Big\{ \langle v | U_{\rm VP} | \xi_v \rangle + \langle c | U_{\rm VP} | \xi_c \rangle \Big\}  \nonumber \\[1mm]
  &&   + \langle cv | I'(\veps_v-\veps_c) | vc \rangle
         \Big\{ \langle c | U_{\rm VP} | c \rangle - \langle v | U_{\rm VP} | v \rangle \Big\} \Big], \label{eq:ScrVP_a}
\end{eqnarray}
where $I_{\rm VP}(\veps)$ is the interelectronic-interaction operator modified by the electron loop
\begin{eqnarray}
I_{\rm VP}(\veps,\bx,\by) &=& \frac{\alpha^2}{2\pi i} \int_{-\infty}^{\infty} \! d\omega  \int \! d\br_1d\br_2 \,
         \frac{\alpha_{1\mu} \exp(i|\veps||\bx-\br_1|)}{|\bx-\br_1|}
         \frac{\alpha_{2\nu} \exp(i|\veps||\by-\br_2|)}{|\by-\br_2|} \nonumber \\[1mm]
  &&  \times  {\rm Tr} \Big[ \alpha^\mu G(\omega-\veps/2, \br_1,\br_2)
                             \alpha^\nu G(\omega+\veps/2, \br_2,\br_1) \Big],
		\label{eq:I_VP}
\end{eqnarray}
and $\veps$ is the energy of the transferred photon.
Finally, for the counterterm VP diagrams~(h) and (j) in Fig.~\ref{fig:qed} one can obtain
\begin{eqnarray}
\Delta E_{\rm \ref{fig:qed}(h)} &=& \langle v | U^{\rm scr}_{\rm VP} | v \rangle,
		\label{eq:ScrVP_CT1} \\[1mm]
\Delta E_{\rm \ref{fig:qed}(j)} &=& 2 \langle v | U_{\rm VP} | \chi_v \rangle,
		\label{eq:ScrVP_CT2}
\end{eqnarray}
where the potential $U^{\rm scr}_{\rm VP}$ corresponding to the vacuum loop with
the additional vertex $\delta V = -V_{\rm scr}$ is defined as follows
\begin{equation}
U^{\rm scr}_{\rm VP}(\bx) =  \frac{\alpha}{2\pi i} \int \! d\br_1 \, \frac{1}{|\bx-\br_1|}
          \int_{-\infty}^\infty \! d\omega \int \! d\br_2 \,
          {\rm Tr} \left[  G(\omega, \br_1,\br_2) \delta V(\br_2) G(\omega, \br_2,\br_1) \right].
		\label{eq:VPpot_scr}
\end{equation}

All the VP contributions are conveniently divided into the Uehling and Wichmann-Kroll parts by
expanding the vacuum-loop electron propagators in powers of the binding potential.
The Uehling parts which correspond to the first nonvanishing terms in these expansions are
ultraviolet divergent. The charge renormalization allows one to find the finite contributions,
for details see Refs.~\cite{Soff:1988:5066,Manakov:1989:1167:note,Persson:1993:2772,
Artemyev:1999:45,Sapirstein:2003:042111}.
For example, the renormalized expression for the Uehling part of the potential~(\ref{eq:VPpot})
induced by the nuclear potential $V_{\rm{nucl}}$ is well-known \cite{Uehling:1935:55,Serber:1935:49}
\begin{eqnarray}
U^{\rm Ue}_{\rm VP}(x)  &=&  -\alpha Z \frac{2}{3}\frac{\alpha}{\pi} \int^\infty_0 \! dr \, 4 \pi r
           \rho_{\rm nucl}(r) \int^\infty_1 \! d t \, \left( 1+\frac{1}{2t^2} \right)
                \frac{\sqrt{t^2-1}}{t^2}  \nonumber \\[1mm]
 && \times \frac{ \{ \exp(-2m|x-r|t)-\exp(-2m(x+r)t) \} }{4mxt},
		\label{eq:Uepot}
\end{eqnarray}
where $\rho_{\rm nucl}$ is the nuclear charge density, normalized to unity:
$\int \!  d\br \, \rho_{\rm nucl}(\br) = 1$.
In order to take into account the screening effects for the Uehling potential
in the diagrams~(f), (i), and (j) one has to replace
$Z\rho_{\rm nucl}$ in Eq.~(\ref{eq:Uepot}) with $Z\rho_{\rm nucl}-N_{\rm scr}\rho_{\rm scr}$.
The charge density $\rho_{\rm scr}$ corresponds to the screening potential $V_{\rm scr}$,
and it is also assumed to be normalized to unity. To obtain the expression for the
Uehling part of the counterterm VP potential~(\ref{eq:VPpot_scr}) one should
replace $Z\rho_{\rm nucl}$ with $N_{\rm scr}\rho_{\rm scr}$ in Eq.~(\ref{eq:Uepot}).
Note, that the Uehling contributions corresponding to the screening potential cancel each other
when the sum of the diagrams~(f) and (h) is considered. However, it is not true for the higher-order
Wichmann-Kroll corrections. The renormalized expression for the Uehling part of the operator~(\ref{eq:I_VP})
has the form
\begin{eqnarray}
I^{\rm Ue}_{\rm VP}(\varepsilon, \bx, \by) \!&=&\!
\alpha \frac{\alpha_{1\mu}\alpha^\mu_2}{|\bx-\by|}
\frac{2}{3}\frac{\alpha}{\pi}
\int^\infty_1 \! dt \, \left( 1+\frac{1}{2t^2} \right) \frac{\sqrt{t^2-1}}{t^2} \nonumber \\[1mm]
&& \times \exp(-\sqrt{(2mt)^2-\varepsilon^2}\,|\bx-\by|).
		\label{eq:UehlOper}
\end{eqnarray}

The Wichmann-Kroll part of the VP potential $U_{\rm VP}$ is given by the expression
\begin{eqnarray}
U^{\rm WK}_{\rm VP}(x) &=& \frac{2\alpha}{\pi} \sum_{\kappa=\pm1}^{\pm\infty} |\kappa|
\int^\infty_0 \!  d\omega \,  \int^\infty_0 \! dr_1 \,  r_1^2 \int^\infty_0 \! dr_2 \, r_2^2
 \frac{1}{{\rm max}(x,r_1)}   V_{\rm eff}(r_2)  \nonumber \\[1mm]
&& \times\sum_{i,k=1}^2 {\rm Re} \Big\{ F^{ik}_\kappa(i\omega,r_1,r_2)[G^{ik}_\kappa(i\omega,r_1,r_2)
                              -F^{ik}_\kappa(i\omega,r_1,r_2)] \Big\},
		\label{eq:WKpot}
\end{eqnarray}
where $G^{ik}_\kappa$ and $F^{ik}_\kappa$ are the radial components of the partial contributions to
the bound- and free-electron Green functions, respectively. In order to obtain the Wichmann-Kroll
part of the counterterm VP potential $U^{\rm scr}_{\rm VP}$ one has to replace
$V_{\rm eff}$ with $\delta V$ and $F^{ik}_\kappa [G^{ik}_\kappa-F^{ik}_\kappa]$ with
$(G^{ik}_\kappa)^2-(F^{ik}_\kappa)^2$.
The Wichmann-Kroll contribution to the formula~(\ref{eq:ScrVP_b})
can be obtained by considering the partial expansion of the difference between the expression~(\ref{eq:I_VP})
and the corresponding equation with the bound-electron Green functions replaced by those for the free electrons.



%


\end{document}